\documentclass{IEEEtran4PSCC}
\ifCLASSINFOpdf
   \usepackage[pdftex]{graphicx}
\else
   \usepackage[dvips]{graphicx}
\fi
\usepackage[cmex10]{amsmath}

\usepackage{graphicx}
\usepackage{float}
\usepackage{amsthm}
\usepackage{amsmath}
\usepackage{amssymb}
\usepackage{commath} 
\usepackage{mathtools}
\usepackage[linesnumbered,ruled,vlined]{algorithm2e}
\usepackage{pifont}
\usepackage{float}
\usepackage{xcolor}
\usepackage{cite}
\usepackage{booktabs}
\usepackage{multirow}
\usepackage{array}
\usepackage{makecell}
\usepackage{siunitx}
\usepackage{rotating}
\usepackage[export]{adjustbox}
\usepackage[nameinlink,capitalize]{cleveref}
\usepackage{parskip}
\usepackage{epstopdf}
\usepackage{url}
\usepackage[caption=false,font=footnotesize]{subfig}

\SetKwInput{KwInput}{Input} 
\SetKwInput{KwOutput}{Output}  
\SetKwInput{KwInitialization}{Initialization}  

\newcommand{\vect}[1]{\boldsymbol{#1}}

\newtheorem{theorem}{Theorem}

\newtheorem*{remark}{Remark}

\theoremstyle{definition}
\newtheorem{definition}{Definition}

\begin{document}
%
\title{Optimal Inverter-Based Resources Placement in Low-Inertia Power Systems}

\author{
\IEEEauthorblockN{Atinuke Ademola-Idowu, Baosen Zhang}
\IEEEauthorblockA{Electrical and Computer Engineering Department\\
University of Washington,\\
Seattle, WA, USA\\
atinukeidowu@gmail.com, zhangbao@uw.edu}
}


\maketitle

\begin{abstract}
An increase in the integration of Inverter Based Resources (IBRs) to the electric grid, will lead to a corresponding decrease in the amount of connected synchronous generators, resulting in a decline in the available rotational inertia system-wide. This can lead to pronounced frequency deviations when there are disturbances and faults in the grid. This decline in available rotational inertia can be compensated for by fast acting IBRs participating in frequency response services, by rapidly injecting into or removing power from the grid. Currently, there are still relative small number of sizable IBRs in the grid. Therefore, the placement of the IBRs in the system, as well as the inverter configuration type and controller, will have a material impact on the frequency response of the grid. 

In this work, we present an optimal placement algorithm that maximizes the benefits of utilizing IBRs in providing frequency response services. That is, we minimize the overall system frequency deviation while using a minimal amount of electric power injection from the IBRs. The proposed algorithm uses the resistance distance to place the IBRs at nodes that are central to the rest of the nodes in the network thus minimizing the distance of power flow. The proposed greedy algorithm achieves a near optimal performance and relies on the supermodularity of the resistance distances. We validate the performance of the placement algorithm on three IEEE test systems of varying sizes, by comparing its performance to an exhaustive search algorithm. We further evaluate the performance of the placed IBRs on an IEEE test system, to determine their impact on frequency stability. The IBRs are configured in a grid-forming mode and equipped with a model predictive control (MPC)-based inverter power control.

\end{abstract}


\thanks{\noindent The authors are partially supported by NSF grants ECCS-1942326.}

\section{Introduction} \label{section:introduction}
The large-scale integration of Inverter Based Resources (IBRs) \cite{wsj.20210516, eia.20191002} will lead to a corresponding decrease in the number of connected synchronous generators, resulting in a decline in the available rotational inertia system-wide. This can potentially result in pronounced frequency deviations from the nominal when there are disturbances and faults in the grid. As a response, system regulators have started mandating that IBRs  participate in providing essential grid services such as frequency response \cite{denholm2020inertia,matevosyan2019evolution}. Due to the fast actuation capability of the IBRs, they can participate in frequency regulation services by rapidly injecting or removing active power to or from the grid~\cite{bevrani2010renewable,ademola2020frequency}. 

An example that demonstrates the potentials of IBRs in grid services is the South Australia's Hornsdale Power Reserve, which utilizes Tesla batteries in providing frequency response services with an estimated cost savings of $\$116$ million in 2019~\cite{esn,aurecon}. The benefits accrued can be further maximized by strategically locating these IBRs in the network. Some IBRs, such as solar and wind, are situated based on availability of resources. Other IBRs, such as energy storage, are more flexible and can either be paired with a solar or wind resource, or can be located strategically to maximum their benefit to the grid where necessary. Optimally placing IBRs in the system can result in a more cost effective and efficient participation in frequency response services by reducing frequency deviations using the minimal overall active power injection from the IBRs into the grid.

Varying techniques have been proposed to solve this problem. These techniques typically rely on treating the frequency dynamics of the IBRs similarly as synchronous machines, in a configuration known as virtual synchronous machines~(VSMs)~\cite{tamrakar2017virtual,AdemolaIdowu2018OptimalDO}. Under this configuration, the optimal IBR placement will be located where the virtual inertia and damping gains of the virtual synchronous machines are maximized~\cite{poolla2017optimal}. The drawback of this approach is that the fast acting capabilities of the IBR are limited to only provided virtual inertia, ignoring that they can provide more adaptive controls~\cite{ademola2020frequency}. In addition, the network is often reduced to nodes containing either machines or IBRs, using the Kron reduction method. This results in only a subset of the nodes being considered for IBR placement, which is suboptimal.


In this paper, we present an optimal IBR placement algorithm that minimizes the overall system frequency deviation while utilizing a minimal amount of power injection from the IBRs, in the event of a disturbance to the network. This algorithm functions by selecting the most central nodes, that is, the nodes that minimize the resistance distance between themselves and the remaining nodes in the network. Under this algorithm, all the nodes in the network can be considered as potential IBR placement points instead of only the nodes with generators. To find the optimal nodes quickly, especially for larger networks, we show that the resistance distance matrix for a power systems network is supermodular. We can therefore adopt a greedy algorithm to determine the optimal nodes efficiently~\cite{nemhauser1978analysis}. We also provide a comparison between the exhaustive search algorithm and our placement algorithm for three different sized power systems. The results show that our algorithm returns similar results to using exhaustive searches, but is orders of magnitude more efficient. 

To validate the optimality of the IBR placement algorithm on system frequency dynamics, we show that the placed IBRs, which are configured in a grid-forming mode, effectively reduce frequency deviations after a fault in the system. The exact frequency control algorithm used for the IBRs is a MPC-based inverter power control algorithm that determines the active power set-point of an IBR~\cite{ademola2020frequency}.

The remainder of this paper is organized as follows: Section \ref{section:structure} defines the models used in this paper. Section \ref{section:algorithm} presents the optimal placement algorithm. Section \ref{section:frequency} presents the inverter power controller used for frequency control of the grid-forming IBRs. Section \ref{section:results} presents the results of the placement algorithm tested on three system of varying sizes namely - IEEE New England 10 machines 39 bus system, IEEE 50 machined 145 bus system and IEEE 69 machines 300 bus system. The frequency response performance is validated on the IEEE New England 10 machines 39 bus system. Section~\ref{section:conclusion} concludes the paper. 





\section{System Structure and Dynamics} \label{section:structure}
We denote the real line by $\mathbb{R}$ and the cardinality of a finite set $\mathcal{S}$ by $\vert \mathcal{S} \vert$. Matrices and vectors are denoted by bold-faced uppercase and lowercase variables, respectively. The $n \times n$ identity matrix is represented as $\vect{I}_n$, the $n \times n$ zero matrix as $\vect{0}_n$, and the $i^{th}$ standard basis vector as $\vect{e}_i$.

\subsection{Frequency Dynamics and Power Flow} \label{subsection:dynamics}
The electric power system is a network of synchronous generators, loads, buses, transmission lines and more recently, IBRs. The frequency of the system is modeled by the rotor dynamics of each synchronous generator, which is governed by the well-known swing equation \cite{sauer2017power}. We use the discretized form of the equations, which in per unit (p.u.) system is:
\begin{equation} \label{eqn:swing_dis}
\begin{aligned}
\omega^{t+1}_i & = \omega^{t}_i + \frac{h}{m_i} \left(P_{\text{m},i}^t - P_{\text{e},i}^t - d_i \omega^{t}_i \right) \; \forall i \in \mathcal{K} \\
\delta^{t+1}_i & = \omega_{\text{b}} \left(\delta^{t}_i + h \ \omega^{t+1}_i\right) \; \forall i \in \mathcal{K},
\end{aligned}
\end{equation}
 where $\mathcal{K}$ is the set of all generators, $h$ is the step size for the discrete simulation, $\delta_i$ (rad) is the rotor angle, $\omega = \bar{\omega}_i - \omega_0$ is the rotor speed deviation, $\omega_{\text{b}}$ is the base speed of the system, $m_i$ is the inertia constant, $d_i$ is the damping constant, $P_{\text{m},i}$ is the mechanical input power and $P_{\text{e},i}$ is the electric power output of the $i^{th}$ machine. 

The electrical output power $P_{\text{e},i}$ is given by the AC power flow equation:
\begin{align} \label{eqn:pf_dis}
P_{\text{e},i}^t = \sum_{i \sim j} \vert E_i E_j \vert [g_{ij} \cos(\delta_i^t - \delta_j^t) + b_{ij}\sin(\delta_i^t - \delta_j^t)],
\end{align}
where $i \sim j$ means all buses $j$ connected to $i$ and
 $g_{ij} + j b_{ij}$ is the admittance between nodes $i$ and $j$. We assume the internal emf are constant because of the actions of the exciter systems.

For placing IBRs, we linearize the AC power flow equations and use the standard DC power flow model. During simulations, we will use the full AC power flow in~\eqref{eqn:pf_dis}. The DC power flow, in vector form, is:
\begin{equation}\label{eqn:pf_dc}
    \textbf{p}^t = \textbf{B} \theta^t, 
 \end{equation}
 where $\textbf{p}$ is a vector of the real power at each bus, $\textbf{B}$ is matrix of admittances between buses, and $\theta$ is the voltage angles referenced to slack.




\subsection{Resistance Distance and Centrality} \label{subsection:distance}
The topology of the electric power grid interconnection can be represented as a weighted undirected graph $\mathcal{G} = (\mathcal{V, E, W})$. Under this graph representation, $\mathcal{V}$, which is the set of nodes represents the electric buses in the grid;  $\mathcal{E}$, which is the set of edges represents the transmission lines which are bidirectional (hence undirected); and  $\mathcal{W}$, which is the set of weights assigned to each edge of the graph, represents the admittance of the transmission lines. Let $n = \vert \mathcal{V} \vert$ represents the number of nodes in $\mathcal{G}$. The Laplacian matrix of $\mathcal{G}$ is a $n \times n$ real symmetric matrix defined as $\textbf{L} = \textbf{D} - \textbf{A}$ where $\textbf{D}$ is the node degree diagonal matrix and $\textbf{A}$ is the weighted adjacency matrix~\cite{baker2006metrized}. 

\begin{definition} For the graph $\mathcal{G}$, the resistance distance \cite{klein1993resistance} between two nodes $i, j \in \mathcal{V}$ is given as:
\begin{equation} \label{eqn:node2node}
\begin{aligned}
    R(i, j) & = (\vect{e}_i - \vect{e}_j)^T \ \textbf{L}^\dagger \ (\vect{e}_i - \vect{e}_j) \\
    & = \textbf{L}_{ii}^\dagger - 2\textbf{L}_{ij}^\dagger + \textbf{L}_{jj}^\dagger,
\end{aligned}
\end{equation} 
where $\textbf{L}^\dagger$ is the Moore-Penrose pseudo-inverse of $\textbf{L}$ and is also a $n \times n$ real symmetric matrix \cite{ranjan2013geometry}. The pseudo-inverse of $\textbf{L}$ is used because $\textbf{L}$ is singular and therefore has no inverse.
\end{definition}
\begin{definition} The resistance distance of a node $i \in \mathcal{V}$ is the sum of the resistance distance between the node $i$ and all other nodes in $\mathcal{V}$, and can be expressed as \cite{bozzo2013resistance}:
\begin{equation} \label{eqn:node2all}
    R(i) = \sum_{j\in V} R(i,j)
        = n \textbf{L}_{ii}^\dagger + \operatorname{Tr}{\left(\textbf{L}^\dagger\right)}
\end{equation} 
\end{definition}

The resistance distance is the distance function on $\mathcal{G}$ \cite{ghosh2008minimizing} and is a measure of the centrality/closeness of the node to other nodes in the network. In the case of an electric power system, a node with a lower resistance distance implies an easier flow of power/current from that node to other nodes. 

\subsection{Supermodularity}
\label{subsection:modularity}
Supermodularity of a function on a discrete set is analogous to the notion of concavity for functions over continuous sets. It characterizes the idea of diminishing returns, where adding an  element to a smaller set gives a larger change in the function than adding the same element to a larger set. A formal definition of this term is given below:
\begin{definition}[Supermodularity]
Let $\mathcal{V}$ be a finite set and $f : 2^\mathcal{V} \to \mathbb{R}$ be a set function on $\mathcal{V}$. The function $f$ is supermodular if $f(\mathcal{S} \cap \mathcal{T}) + f(\mathcal{S} \cup \mathcal{T}) \geq f(\mathcal{S}) + f(\mathcal{T})$ or $f(S) - f (S \cup \{u\}) \geq f(T) - f (T \cup \{u\})$, for any subsets $\mathcal{S} \subseteq \mathcal{T} \subseteq \mathcal{V}$, and any element $u \in \mathcal{V} \backslash \mathcal{T}$.
\end{definition}
If a function over a set is supermodular, then a greedy algorithm can be used to solve the problem efficiently. The greedy algorithm has been proven in ~\cite{nemhauser1978analysis, cornuejols1977exceptional} to have a polynomial-time complexity and performance bound within a constant of the optimum. It should be noted that if the function $f$ is supermodular, then $-f$ is submodular, and vice versa.


In addition to supermodularity, we will use the notion of a non-increasing function over a finite set, defined as: 
\begin{definition}[Monotonicity]
The set function $f : 2^\mathcal{V} \to \mathbb{R}$ is monotonously non-increasing if $f(\mathcal{T}) \leq f(\mathcal{S})$, for all $\mathcal{S} \subseteq \mathcal{T}$.
\end{definition}

The concept of set ordering between sets can be used to show the relational structure between sets and can be defined as: 
\begin{definition}[Ordered Set]
A set $\mathcal{S}$ is linearly ordered if the relation $\leq$ on $\mathcal{S}$ for all $s,t,u \in \mathcal{S}$ satisfies the properties of reflexivity $s \leq s$, anti-symmetry (if $s \leq t$ and $t \leq s$, then $s = t$), transitivity (if $s \leq t$ and $t \leq u$, then $s \leq u$), and Trichotomy law (either $s \leq t$ or $t \leq s$).
\end{definition}

\section{Optimal Placement Algorithm} \label{section:algorithm}
\subsection{Problem Formulation} \label{subsection:formulation}
The objective of this work is to determine the best location in the grid to place a specified number of IBRs, such that when there is a disturbance to the grid, these IBRs can participate in efficiently stabilizing the grid frequency. To achieve this, we will capitalize on the relationship between frequency and electric power flow in the grid.

From the swing equation in \eqref{eqn:swing_dis}, it can be observed that the rotor speed deviation and as such the frequency deviation is proportional to the power imbalance. This implies that frequency deviations and rate of change of frequency (ROCOF) can be curtailed by minimizing the power imbalance. The power imbalance can be minimized by either increasing or reducing the the power generation or power consumption. One of the advantages of IBRs is their fast actuation capabilities which can enable them participate in frequency control by rapidly injecting or absorbing active power in the grid. Since the disturbance to the grid can occur at any location, typically unknown beforehand, the impact of the IBRs on frequency control can be maximized by strategically locating them at the ``central'' nodes in the system.

Equation \eqref{eqn:pf_dc} shows that the power flow in an electric grid is dependent on the topology of the grid, that is, dependent on the susceptance matrix $\vect{B}$. 
This matrix is a graph Laplacian and we can define the ``resistance'' between two buses through \eqref{eqn:node2node}. Of course, since we work with the DC power flow, there are no losses and these are not the actual resistance on a line. Rather, they serve as a distance measure as defined in Section~\ref{subsection:distance}. For consistent terminology, we still refer to the quantities computed from $\vect{B}$ using \eqref{eqn:node2node} and \eqref{eqn:node2all} as resistances. 

Let the set of nodes containing IBRs be denoted as $\mathcal{I}$ and the remaining nodes in the network as $\mathcal{J}$, such that $\mathcal{V} := \mathcal{I} \cup \mathcal{J}$. The placement problem can be stated as selecting $k$ number of nodes to place the IBRs that minimizes the resistance distance $R(\mathcal{I})$ between nodes with IBRs and nodes without IBRs.
Mathematically, this can be written as:
\begin{equation} \label{eqn:placement}
\begin{aligned}
\underset{\mathcal{I} \in \mathcal{V} }{\text{min}}& \quad R(\mathcal{I})   =  \sum_{j \in \mathcal{J}}  \underset{i \in \mathcal{I}}{\text{min}} \quad R(i, j )\\
s.t.& \quad  \vert \mathcal{I}\vert = k
\end{aligned}
\end{equation}
For each node $i \in \mathcal{I}$ and $j \in \mathcal{J}$, $R(i, j )$ is the resistance distance between node $i$ and $j$ as defined in \eqref{eqn:node2node}.

\subsection{Supermodularity of Distance Function} \label{subsection:distancemodularity}
In order to solve the optimization problem posed in \eqref{eqn:placement} efficiently, we can take advantage of the properties of the resistance distance function $R$. 
\begin{theorem} \label{thrm:mono}
The resistance distance function  $R(\mathcal{S})$ is a monotonically non-increasing function of the set of vertices $\mathcal{S}$, such that, $R(\mathcal{T}) \leq R(\mathcal{S})$ for any subsets $\mathcal{S} \subseteq \mathcal{T} \subseteq \mathcal{V}$.
\end{theorem}
\begin{proof} \renewcommand{\qedsymbol}{}
Let $\mathcal{S}$ and $\mathcal{T}$ be linearly ordered sets and $\mathcal{Q}\coloneqq \mathcal{T} \backslash \mathcal{S}$ be defined as a linearly ordered set such that $\mathcal{T} = \mathcal{S} \cup \mathcal{Q}$ and $\mathcal{Q}_m = \{q_1, q_2, \dots q_m\}$ for $m = \vert\mathcal{T}\vert - \vert\mathcal{S}\vert$. From the definition of resistance distance, we have that:
\begin{equation*}
\begin{aligned}
    R(\mathcal{S} \cup Q_1) = 
    R(\mathcal{S} \cup \{q_1\})
    & = \sum_{j \in \mathcal{J}} \text{min}  \{R(\mathcal{S}, j), R(q_1, j)\} \\
    & \leq \sum_{j \in \mathcal{J}} R(\mathcal{S}, j) = R(\mathcal{S}) \\
\end{aligned}
\end{equation*}
Adding the elements of set $\mathcal{Q}$ sequentially to set $\mathcal{S}$ up to the set $\mathcal{T}$ to obtain
\begin{equation*}
\begin{aligned}
    R(\mathcal{T}) = 
    R(\mathcal{S} \cup Q_m) & = 
    R(\mathcal{S} \cup \{q_1, \dots q_m\})  \\
    = \sum_{j \in \mathcal{J}} & \text{min} \{R(\mathcal{S}, j), R(q_1, j) \dots R(q_m, j)\} \\
    & \leq \sum_{j \in \mathcal{J}} R(\mathcal{S}, j) = R(\mathcal{S})
    \qedhere
\end{aligned}
\end{equation*}
\end{proof}
Intuitively, this means that as the number of selected nodes increases, the resistance distance between those nodes and the remaining nodes in the network decreases.
\begin{theorem} \label{thm:super}
If the the resistance distance function  $R(\mathcal{S})$ is a monotonically non-increasing function of the set of vertices $\mathcal{S}$, then $R(\mathcal{S})$ is supermodular, thus 
$$R(\mathcal{S} \cap \mathcal{T}) + R(\mathcal{S} \cup \mathcal{T}) \geq R(\mathcal{S}) + R(\mathcal{T})$$
 for any subsets $\mathcal{S} \subseteq \mathcal{T} \subseteq \mathcal{V}$.
\end{theorem}
\begin{proof}\renewcommand{\qedsymbol}{}
Let $\mathcal{S}$ and $\mathcal{T}$ be linearly ordered sets and $\mathcal{Q}\coloneqq \mathcal{T} \backslash \mathcal{S}$ be defined as a linearly ordered set such that $\mathcal{Q} = \mathcal{S} \cap \mathcal{T}$. We have that:
\begin{equation*}
\begin{aligned}
R(\mathcal{S} \cap \mathcal{T}) + R(\mathcal{S} \cup \mathcal{T}) & = 
R(\mathcal{S} \cap \mathcal{T}) + R(\mathcal{T}) \\
& = R(\mathcal{Q}) + R(\mathcal{T})\\
& \geq  R(\mathcal{S}) + R(\mathcal{T}) 
\qedhere
\end{aligned}
\end{equation*}
\end{proof}
The inequality on the last line follows from Theorem \ref{thrm:mono} since $S$ is a linearly ordered set such that $R(\mathcal{Q}) \geq R(\mathcal{S})$.
\begin{remark}
Results similar to those in Theorems~\ref{thrm:mono} and~\ref{thm:super} have appeared in the literature, for example, see~\cite{shan2018improving,li2019current}. However, those results are often stated in complicated terminology and the proofs are complex. The proof presented here are considerably shorter and more elementary. 
\end{remark}

\subsection{Greedy Algorithm} \label{subsection:algorithm}
With $R(\mathcal{I})$ proven to be a supermodular set function, equation \eqref{eqn:placement} is the minimization of a supermodular function with cardinality constraint. This is equivalent to the maximization of a submodular function with cardinality constraint \cite{contreras2014hub, wolsey1982maximising} and has been proven to be NP-hard \cite{nemhauser1978analysis}. Therefore an optimal algorithm will be a version of exhaustive search, which will be prohibitively expensive.

However, efficient algorithms that are suboptimal have been established. For example, the simple greedy algorithm proposed in \cite{nemhauser1978analysis,wolsey1982maximising} will find solutions that are within $37\%$ of the optimal solution. Using this algorithm, let the optimal set of IBR placements be $\mathcal{I}^*$ and the number of IBRs to be placed be $k$. The optimal node $i^*$ at each iteration is the node which when added to $\mathcal{I}^*$, that is $R(\mathcal{I}^* \cup \{i^*\},j)$ minimizes the resistance distance between $\mathcal{I}^* \cup \{i^*\}$ and $j \in \mathcal{V}\backslash\mathcal{I}$. The pseudocode of this algorithm is shown in Algorithm \ref{alg:greedy}.




\begin{algorithm}
\caption{IBR Placement Algorithm}\label{alg:greedy}
\DontPrintSemicolon
\KwInput{Graph $\mathcal{G} = (\mathcal{V, E, W})$\; \quad \qquad  No. of IBRs $k$}
\KwOutput{Optimal set of IBR locations $\mathcal{I}^*$}
\KwInitialization{$\mathcal{I}^* \leftarrow \emptyset$}
Compute $R(i,j)$ using \eqref{eqn:node2node}\;
\While{$\vert \mathcal{I}^* \vert < k$}{
   		$i \leftarrow \underset{j\in \mathcal{V}\backslash\mathcal{I} }{\text{argmin}}
   		 \ \ R(\mathcal{I}^*,j)$ \; 
   		$\mathcal{I}^* \leftarrow \mathcal{I}^* \cup \{i\}$\;
   }
\Return{$\mathcal{I}^*$} 
\end{algorithm}

\section{Frequency Control in Low-Inertia Systems} \label{section:frequency}
After selecting the optimal location to place the IBRs using algorithm \ref{alg:greedy}, these IBRs need to be controlled to enable them participate effectively in providing frequency control in a low-inertia power system. Varying control strategies for the IBRs have been proposed (e.g., see \cite{milano2018foundations, tamrakar2017virtual,AdemolaIdowu2018OptimalDO} and the references within). A fundamental drawback of these control strategies are the constraining of IBRs to behave like synchronous machines when responding to frequency events, thus limiting the potentials of the fast acting and flexible IBRs. To address this, we utilize the MPC-based inverter power controller algorithm developed in \cite{ademola2020frequency}. Based on this controller's mode of operation, it can be categorized as a power controller for a grid-forming inverter's power control loop. The controller functions by changing the active power set-point of the IBRs, which are configured in a grid-forming mode, based on frequency measurements to counterbalance the power imbalance, as a result of disturbances to the system.

An optimization problem is formulated to determined this
active power set-point of the IBRs by using the IBR angle as a control variable, such that the frequency control objectives are satisfied. Mathematically, this can be written as follows: 
\begin{subequations} \label{eqn:opt_alg}
\begin{align} 
    & \underset{\{\vect{u}^0, \vect{u}^1, \dots, \vect{u}^{N-1}\}}{\text{min}} \ \sum_{t=0}^{N-1}
    \label{eqn:obj}
    \left\{\Vert \vect{\omega}^{t+1} \Vert^2_2 \ + \frac{1}{h}\Vert \vect{\omega}^{t+1} - \vect{\omega}^t \Vert^2_2 \right. \\
    & \qquad \qquad \qquad \qquad \quad \left. + \Vert \vect{r} \odot \vect{P}_\text{ibr}^t \Vert^2_2 \right\}\\
    & \text{s.t.}
    \label{eqn:speed}
    \ \omega^{t+1}_i= \omega^t_i + \frac{h}{m_i} \big(P_{\text{m},i}^t - P_{\text{e},i}^t  - d_i \omega^t_i  - \bigtriangleup P_i^t \big), \ \forall i \in \mathcal{G}\\
    \label{eqn:Pe}
    & P_{e,i}^t = \text{Equation} \ \eqref{eqn:pf_dis}, \ \forall i \in \mathcal{G}\\
    \label{eqn:Pinv}
    & P_{ibr,k}^t = \text{Equation} \ \eqref{eqn:pf_dis}, \ \forall k \in \mathcal{I},
\end{align}
\end{subequations}
where $\vect{u}^t \in \mathcal{R}^{\mathcal{|I|}}$ is the vector of all IBR angles (referenced to the slack-bus) and represents the control variable in the optimization problem. $\vect{\omega}^{t+1} \in \mathcal{R}^{\mathcal{|G|}}$ is a vector of all machine frequency deviations at the next time step, $\vect{\omega}^{t+1} - \vect{\omega}^t$ is a vector of all machine ROCOF between the current and next time step. The evolution of $\vect{\omega}$ is given in \eqref{eqn:speed} (swing equations) with the added $\bigtriangleup P_i$ used to denote disturbances to the network which can be either a loss in generation or load. $\vect{P}_\text{e}^t \in \mathcal{R}^{\mathcal{|G|}}$ and $\vect{P}_\text{ibr}^t \in \mathcal{R}^{\mathcal{|I|}}$ are vectors of all generator and IBR output power respectively. Depending on how the system is interconnected, $\vect{P}_\text{ibr}^t$ can also be written in the form of \eqref{eqn:pf_dis} \cite{ademola2020frequency}. The IBR droop $\vect{r} \in \mathcal{R}^{\mathcal{|I|}}$ is a vector of weights for the IBRs output power and $\odot$ is the componentwise product between two vectors. 

This optimization problem in \eqref{eqn:opt_alg} can be reduced to a linear quadratic programming (LQR) problem by using the DC power flow representation in \eqref{eqn:pf_dc} to isolate the control variable (IBR angle), converting \eqref{eqn:swing_dis} to state space form, and integrating an observer for disturbance estimation. The final LQR form is:
\begin{equation} \label{eqn:opt_inv}
\begin{aligned}
& \underset{\vect{u^t}}{\text{min}} \
J = \frac{1}{2} \sum_{t = 0}^{N-1} \left[ \vect{y^{t^T} Q_1 y^t} + \vect{\bigtriangleup y^{t^T} Q_2 \bigtriangleup y^t} \right]\\ 
& \text{s.t.}
\quad \vect{z^{t+1} = A z^t +  B u^t}\\
& \qquad \quad \  \vect{y^t  = C z^t},
\end{aligned}
\end{equation}

can be solved as an infinite-horizon problem, where $z$ is the a vector of system states and disturbances. The optimal solution is linear in the starting point $\vect{z}^0$ such that $\vect{u^*} = \vect{-H^{-1} F^T z^0}$. Details on the structure of the matrices $\vect{A}, \vect{B}, \vect{C}, \vect{H}$ and $\vect{F}$ can be found in \cite{ademola2020frequency}. It should be noted that the actual control of the IBR is not done via angle control, rather, we use the optimized $u_t$ for every step to find the corresponding active power output of the inverter using \eqref{eqn:pf_dis}, then set the IBRs active power set-point to that value. 

\section{Case Studies} \label{section:results}
\subsection{Optimal Placement} \label{subsection:result_placement}
The efficacy of the placement algorithm in section \ref{section:algorithm} is validated on three test systems of varying sizes - IEEE New England 10 machine 39 bus (IEEE39), IEEE 50 machine 145 Bus System (IEEE145), and IEEE  69 machine  300 Bus System (IEEE300) \cite{zimmerman2010matpower,chow1992toolbox}. For each test system, the specified number of IBRs to be placed ranges from one IBR to four IBRs. The goal of the algorithm is to determine the optimal nodes at which to place the specified amount of IBRs, based on its resistance distance to other nodes in the network. 
\begin{table*}[htbp]
  \centering
  \caption{Placement Algorithm Results for the IEEE NE 10 Machine 39 Bus System (IEEE39)}
  \begin{tabular}{|l|c|S|c|S|c|S|c|S|}
    \hline
     \multirow{2}{*}{\thead{\textbf{Method}}} &
      \multicolumn{2}{c|}{\thead{\textbf{1 IBR}}} &
      \multicolumn{2}{c|}{\thead{\textbf{2 IBRs}}} &
      \multicolumn{2}{c|}{\thead{\textbf{3 IBRs}}} &
      \multicolumn{2}{c|}{\thead{\textbf{4 IBRs}} }\\
      \cline{2-9}
      & {\thead{Bus}} & {\thead{Time (s)}} & {\thead{Bus}} & {\thead{Time (s)}} & {\thead{Bus}} & {\thead{Time (s)}}  & {\thead{Bus}} & {\thead{Time (s)}} \\
      \hline
    {\thead{\textbf{Exhaustive Search}}}  & 16 & 0.04 & 16, 6 & 0.25 & 16, 6, 29 & 1.20 & 16, 6, 29, 2 & 7.69 \\
    \hline
    {\thead{\textbf{Greedy Algorithm}}} & 16 & 0.02 & 16, 6 & 0.04 & 16, 6, 29 & 0.05 & 16, 6, 29, 2 & 0.07\\
    \hline
  \end{tabular}
  \label{table:bus39}
\end{table*}

\begin{table*}[htbp]
  \centering
  \caption{Placement Algorithm Results for the IEEE 50 Machine 145 Bus System (IEEE145)}
  \begin{tabular}{|l|c|S|c|S|c|S|c|S|}
    \hline
    \multirow{2}{*}{\thead{\textbf{Method}}} &
    \multicolumn{2}{c|}{\thead{\textbf{1 IBR}}} &
    \multicolumn{2}{c|}{\thead{\textbf{2 IBRs}}} &
    \multicolumn{2}{c|}{\thead{\textbf{3 IBRs}}} &
    \multicolumn{2}{c|}{\thead{\textbf{4 IBRs}} }\\
      \cline{2-9}
      & {\thead{Bus}} & {\thead{Time (s)}} & {\thead{Bus}} & {\thead{Time (s)}} & {\thead{Bus}} & {\thead{Time (s)}}  & {\thead{Bus}} & {\thead{Time (s)}} \\
      \hline
    {\thead{\textbf{Exhaustive Search}}}  & 12 & 0.08 & 12, 68 & 4.10 & 12, 68, 94 & 171.30 & 12, 68, 94, 142 & 6426.38 \\
    \hline
    {\thead{\textbf{Greedy Algorithm}}} & 12 & 0.04 & 12, 68 & 0.23 & 12, 68, 94 & 0.43 & 12, 68, 94, 142 & 0.52\\
    \hline
  \end{tabular}
  \label{table:bus145}
\end{table*}

\begin{table*}[htbp]
  \centering
  \caption{Placement Algorithm Results for the IEEE 69 Machine 300 Bus System (IEEE300)}
  \begin{tabular}{|l|c|S|c|S|c|S|c|S|}
    \hline
    \multirow{2}{*}{\thead{\textbf{Method}}} &
    \multicolumn{2}{c|}{\thead{\textbf{1 IBR}}} &
    \multicolumn{2}{c|}{\thead{\textbf{2 IBRs}}} &
    \multicolumn{2}{c|}{\thead{\textbf{3 IBRs}}} &
    \multicolumn{2}{c|}{\thead{\textbf{4 IBRs}} }\\
      \cline{2-9}
      & {\thead{Bus}} & {\thead{Time (s)}} & {\thead{Bus}} & {\thead{Time (s)}} & {\thead{Bus}} & {\thead{Time (s)}}  & {\thead{Bus}} & {\thead{Time (s)}} \\
      \hline
    {\thead{\textbf{Exhaustive Search}}} & 245 & 0.20 & 245, 276 & 45.78 & 245, 276, 289 & 3714.77 &  245, 276, 289, 281 &  1056470.74\\
    \hline
    {\thead{\textbf{Greedy Algorithm}}} & 245 & 0.17 & 245, 276 & 0.64 & 245, 276, 289 & 0.95 & 245, 276, 289, 281 &  1.32\\
    \hline
  \end{tabular}
  \label{table:bus300}
\end{table*}
The performance of this placement algorithm is compared to the exhaustive search algorithm (ESA), which tries out every combination of nodes in the network to determine the optimal node placement. The performance metrics will be based on the accuracy in determining the optimal nodes and computation speed. The simulations were carried out using MacBook Pro 2.7 GHz Dual-Core Intel Core i5 and MATLAB version 2019b.

Table \ref{table:bus39} shows the placement results of 1 to 4 IBRs in the IEEE39 system which has 39 nodes. It can be observed that the placement algorithm has a 100\% accuracy in determining the optimal nodes at which to place the IBRs when compared to the ESA but it accomplishes this at a fraction of the time it takes the ESA. The computational time difference is particularly noticeable as the number of IBRs to be placed increases. Compared to the ESA, there is a time saving of $0.02 \ \text{s}$, $0.21 \ \text{s}$, $1.15 \ \text{s}$, and, $7.62 \ \text{s}$ when placing 1, 2, 3, and 4 IBRs respectively. 

The placement algorithm also scales well to larger systems with a higher amount of nodes. Table \ref{table:bus145} shows the placement results of 1 to 4 IBRs in the IEEE145 system
which has 145 nodes. It can be observed that using the placement algorithm yields the same level of accuracy as the ESA when selecting the optimal nodes, but requires a significantly smaller amount of computational time to accomplish this. This shows that an increase in the number of nodes has a minimal impact on the computational time when using the placement algorithm. Compared to the ESA, the time savings accrued when using the placement algorithm is $0.04 \ \text{s}$, $3.87 \ \text{s}$, $170.87 \ \text{s}$, and, $6425.86 \ \text{s}$ when placing 1, 2, 3, and 4 IBRs respectively. 

The same observations can be made on an even larger system as shown in Table \ref{table:bus300}, which shows the placement result of 1 to 4 IBRs in the IEEE300 system
which has 300 nodes. The benefits of the placement algorithm is particularly evident as the size of the system increases. In this case, the number of nodes in the IEEE300 system is about twice the number of nodes in the IEEE145 system, but requires about $165$ times the computational time when using the ESA to place 4 IBRs. This is in comparison to our placement algorithm which utilizes about $0.87$ times the computational power for the same task. That is, to optimally place 4 IBRs, it takes the placement algorithm $1.32 \ \text{s}$ while it takes the ESA about 2 weeks.

From Table \ref{table:bus39} - \ref{table:bus300}, we can conclude that, as the number of IBRs and the system size increases, the computation time increases exponentially for the ESA while it increases linearly for the placement algorithm, without any loss in accuracy. This validates the effectiveness of utilizing the placement algorithm especially for a larger sized system. 

\subsection{Frequency Response} \label{subsection:result_frequency}
\begin{figure}[htbp]
  \centering
  \includegraphics[scale = 0.7]{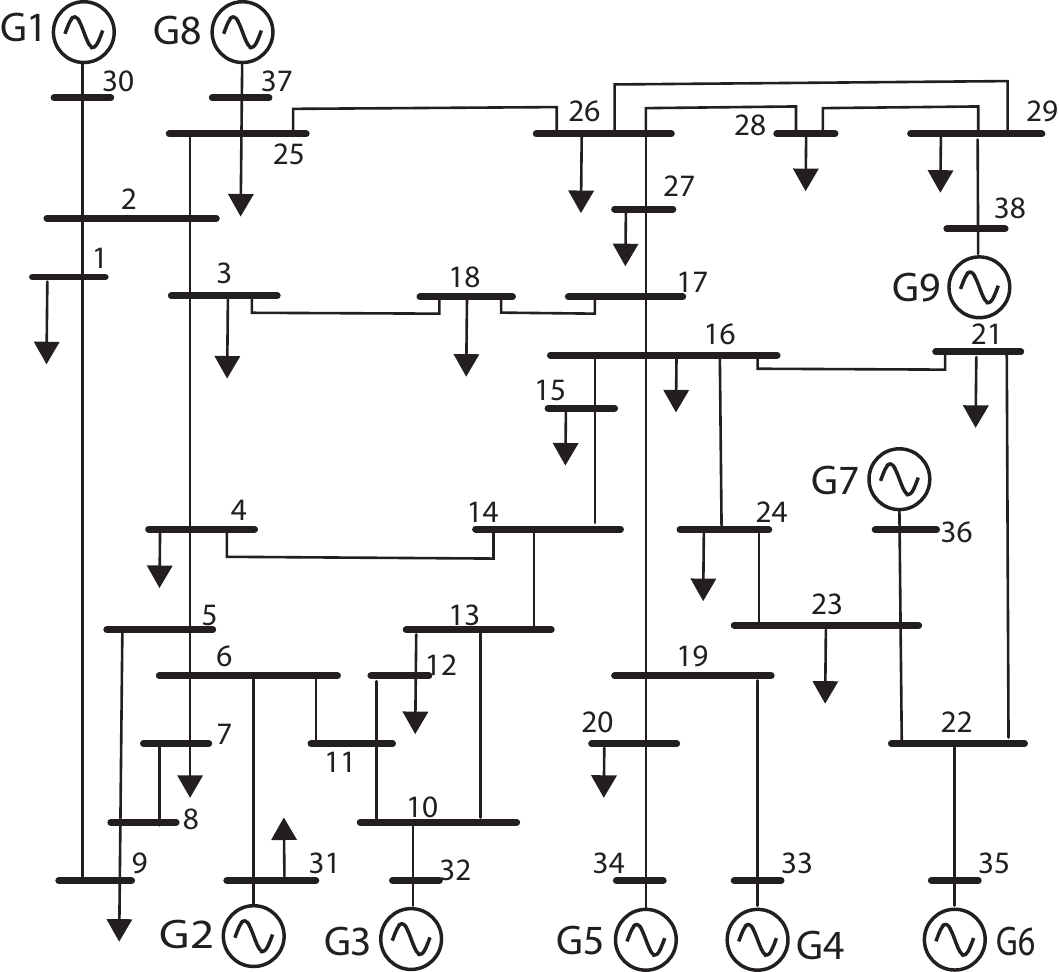}
  \caption{New England 39-bus system schematic. }
  \label{fig:ne39_network}
\end{figure}

The eventual goal of optimally placing the IBRs is to enable the them effectively participate in providing frequency response services in the system. We test the frequency response performance by placing 2 grid-forming IBRs, equipped with the MPC-based inverter power controller discussed in \ref{section:frequency}, in the IEEE39 system described in \ref{subsection:result_placement} and shown in Fig. \ref{fig:ne39_network}. 

\begin{table*}
  \centering
  \caption{Frequency Response Performance Metrics of the Optimal, Next Optimal and Random Bus IBR Placement. $\Vert \textbf{f} \Vert_{1,h}$ is the time-step scaled $L_1$ norm of generator frequency, $\Vert \textbf{P}_{gen} \Vert_{1,h}$ is the time-step scaled $L_1$ norm of generator power and $\Vert \textbf{P}_{ibr} \Vert_{1,h}$ is the time-step scaled $L_1$ norm of IBR power}
   \begin{tabular}{|l|S|S|S|S|S|S|S|S|S|S|S|S|}
    \hline
    \multirow{2}{*}{\thead{\textbf{Disturbed} \\ \textbf{Generator}}} &
      \multicolumn{3}{c|}{\thead{\textbf{Optimal Bus}\\ (Bus 6 and 16)}} &
      \multicolumn{3}{c|}{ \thead{\textbf{Next Optimal Bus} \\ (Bus 5 and 16)}}&
       \multicolumn{3}{c|}{\thead{\textbf{Random Bus} \\ (Bus 3 and 23)}} \\
      \cline{2-10}
      & {\thead{$\Vert \textbf{f} \Vert_{1,h}$ \\ (Hz)}} &  {\thead{$\Vert \textbf{P}_{gen} \Vert_{1,h}$ \\ (p.u.)}} & {\thead{$\Vert \textbf{P}_{ibr} \Vert_{1,h}$ \\ (p.u.)}}  & {\thead{$\Vert \textbf{f} \Vert_{1,h}$ \\ (Hz)}} &  {\thead{$\Vert \textbf{P}_{gen} \Vert_{1,h}$ \\ (p.u.)}} & {\thead{$\Vert \textbf{P}_{ibr} \Vert_{1,h}$ \\ (p.u.)}} & 	{\thead{$\Vert \textbf{f} \Vert_{1,h}$ \\ (Hz)}} &  {\thead{$\Vert \textbf{P}_{gen} \Vert_{1,h}$ \\ (p.u.)}} & {\thead{$\Vert \textbf{P}_{ibr} \Vert_{1,h}$ \\ (p.u.)}}\\
      \hline
    {\thead{\textbf{Gen 1}}} & 0.10 & 60.14 & 47.16 & 0.10 & 63.51 & 50.50 & 0.97 & 118.60 & 106.69 \\
    \hline
    {\thead{\textbf{Gen 2}}} & 1.15 & 79.59 & 58.37 & 1.23 & 83.29 & 61.49 & 2.59 & 135.75 & 108.25 \\
    \hline
    {\thead{\textbf{Gen 4}}} & 2.05 & 83.13 & 54.13 & 2.05 & 86.37 & 57.40 & 2.83 & 138.82 & 106.60 \\
    \hline
    {\thead{\textbf{Gen 6}}} & 2.33 & 87.93 & 56.10 & 2.31 & 91.16 & 59.46 & 1.85 & 135.40 & 109.59\\
    \hline
    {\thead{\textbf{Gen 7}}} & 2.38 & 83.65 & 53.24 & 2.37 & 86.88 & 56.56 & 1.76 & 134.91 & 109.52 \\
    \hline
    {\thead{\textbf{Gen 8}}} & 2.59 & 80.88 & 49.28 & 2.59 & 84.31 & 52.72 & 2.50 & 141.73 & 111.24 \\
    \hline
    {\thead{\textbf{Gen 9}}} & 3.33 & 93.16 & 54.87 & 3.33 & 96.18 & 57.93 & 2.95 & 146.81 & 113.68 \\
    \hline
    {\thead{\textbf{Total} }}& {\thead{\textbf{14.83}}} & {\thead{\textbf{568.48}}} & {\thead{\textbf{373.14}}} & {\thead{\textbf{14.89}}} & {\thead{\textbf{591.70}}} & {\thead{\textbf{396.05}}} & {\thead{\textbf{15.44}}} & {\thead{\textbf{952.03}}} & {\thead{\textbf{765.58}}} \\
    \hline
  \end{tabular}
  \label{table:ne39freq}
\end{table*}

The IEEE39 system is first transformed into a low-inertia system by removing the interconnection to the rest of the US network at bus 1. We assume that the generators 3 and 5 at buses 32 and 34 respectively in Fig. \ref{fig:ne39_network}, are replaced with grid-forming IBRs, (either solar or wind but coupled with energy storage) which has a total aggregated capacity equal to the replaced generator. The presence of a coupled energy storage is to ensure that there would be enough power available for frequency response. The results of optimally placing 2 IBRs in the IEEE39 system is detailed in table \ref{table:bus39} where the optimal nodes are nodes $6$ and $16$. The IBRs located at buses $32$ and $34$ are then relocated to buses $6$ and $16$. After this placement, the rest of the system can then be reduced to an equivalent network using Kron reduction \cite{nishikawa2015comparative}. It should be noted that the generators in the network are equipped with droop and automatic governor control to enable them also respond to frequency events.

The total simulation duration is for $30 \ \text{s}$ and a disturbance in the form of a partial generating capacity loss ($60\%$ loss of capacity) is applied to all the generator in the network one at a time, from from $0.5 \ \text{s}$ to $5 \ \text{s}$, to initiate an event that can lead to a marked frequency decline. This will allow for testing the frequency response impact of the IBRs placement for all possible disturbance locations. The IBRs will be configured in a grid-forming mode and controlled using the algorithm in \ref{section:frequency}. This controller will determine the optimal amount of power output for each IBR such that the frequency deviation and ROCOF is minimized. 
The performance metrics that will be used is the time-step scaled $L_1$ norm and is defined for a parameter \textbf{x} as $\Vert \textbf{x} \Vert_{1,h} = h \cdot \sum_{n=1}^{N} \sum_{t=1}^{T} \vert x^t - x^0 \vert $, where $x^t$ is the parameter value at time $t$, $x^0$ is the nominal value at time $t=0$ and $h$ is the simulation time step. This performance metrics is evaluated on the generator frequency $\Vert \textbf{f} \Vert_{1,h}$, generator power $\Vert \textbf{P}_{gen} \Vert_{1,h}$ and IBR power $\Vert \textbf{P}_{ibr} \Vert_{1,h}$
The overall performance metrics is the total over all the individual metrics for each disturbed generator. The total overall performance is used because for each individual case, the performance might be better due to proximity to the disturbed generator but worse when far away. For best performance, we expect the optimal placement to have the smallest overall performance metrics value. It should be noted that having a minimal IBR power deviation implies that the IBRs utilize a minimum amount of power to restore the frequency to nominal.

The frequency response performance of the optimal node placement is compared to the next optimal node placement as determined by the placement algorithm, in this case, nodes $5$ and $16$, and a random node selection placement, chosen as node $3$ and $23$. Table \ref{table:ne39freq} shows the performance metrics of frequency response of the optimal, next optimal and random bus placement of the IBRs. The overall $\Vert \textbf{f} \Vert_{1,h}$ is $14.83$ Hz, $14.89$ Hz, and $15.44$ Hz, for the optimal, next optimal and random nodes respectively. It can be observed that placing the IBRs at the optimal node as determined by the placement algorithm results in the least overall $\Vert \textbf{f} \Vert_{1,h}$ while randomly selecting a node for placement results in the worst overall $\Vert \textbf{f} \Vert_{1,h}$. The same observations can be made for the total power deviations (both generators and IBRs). It can be noticed that even placing the IBRs at the next optimal location requires utilizing more IBR power (an additional $22.91$ p.u.) and a $0.06$ Hz frequency deviation difference, from the optimal to obtain the same result. In general, the optimal node consistently utilizes a minimum amount of power from the IBR to restore the frequency to nominal. Therefore placing IBRs optimally can result in a lot of savings and better performance in providing frequency response services.


\section{Conclusion} \label{section:conclusion}
In this paper, we proposed an optimal IBR placement algorithm to place IBRs in an electric grid to enable the IBRs participate effectively in frequency regulation services. The algorithm selects as placement nodes, the nodes most central to other nodes in the system using the resistance distance function. The proposed algorithm relies on the supermodularity of the resistance distance function to enable computation within a limited time by using a greedy algorithm. We show via simulation on varying system sizes, the efficacy and time saving benefits of our algorithm. We also show via simulation on a test system, the frequency response benefits of placing the grid-forming IBRs, using the proposed algorithm compared to arbitrarily placing the IBRs in the network. For future works, we will explore the properties of the power system network that results in a tight optimality gap for the greedy algorithm. We will also extend the algorithm validation to power systems with a larger number of nodes.
\bibliographystyle{IEEEtran}
\bibliography{Placement}

\end{document}